\documentclass[prl,aps,twocolumn,groupaddress,longbibliography]{revtex4-1}
\usepackage{nopageno}
\usepackage{latexsym}
\usepackage{lineno}
\usepackage{hyperref}
\usepackage{url}
\usepackage{graphicx}
\usepackage{verbatim}
\usepackage{multirow}
\usepackage{amsmath, bm}
\newcommand{\beq}{\begin{eqnarray}}
\newcommand{\eeq}{\end{eqnarray}}
\usepackage{mathrsfs}
\usepackage{float}
\usepackage[usenames, dvipsnames]{color}
\usepackage{mathtools}
\usepackage{slashed}
\usepackage{physics}	% installed packages for typing maths and physics
\usepackage{graphicx}   % need for Fig.ures
\usepackage{epstopdf}
\usepackage{subfigure}  % use for side-by-side Fig.ures
\usepackage{hyperref}   % use for hypertext links, including those to external documents and URLs
\usepackage{bbold}
\usepackage{wasysym}
\usepackage{amssymb}
\usepackage{feynmp}
\usepackage[symbol]{footmisc}
\def \bs{\textbf}
\usepackage[latin1]{inputenc}
\usepackage{tikz}
\usetikzlibrary{decorations.pathmorphing}
\usetikzlibrary{shapes.misc}
\tikzset{cross/.style={cross out, draw=black, minimum size=8*(#1-\pgflinewidth), inner sep=0pt, outer sep=0pt},
cross/.default={1pt}}
\usetikzlibrary{patterns,math}
\newcommand{\RN}[1]{%
  \textup{\uppercase\expandafter{\romannumeral#1}}%
}

\newcommand{\qs}[1]{{\color{black} #1}}

\begin{document}

\title{Electron correlations and  $T$-breaking density wave order in a $\mathbb{Z}_2$ kagome metal}
\author{Chandan Setty$^{\oplus,\dagger}$}
\author{Haoyu Hu$^{\oplus}$}
\author{Lei Chen}
 \author{Qimiao Si}
 \affiliation{%
Department of Physics \& Astronomy,
Rice Center for Quantum Materials,
Rice University, Houston, Texas 77005,USA
 }%

\begin{abstract}
There have been extensive recent developments on kagome metals, such as
T$_m$X$_n$ (T= Fe, Co and X= Sn, Ge) and $A$V$_3$Sb$_5$ ($A=$ Cs, K, Rb).
An emerging issue is the nature of correlated phases when topologically \textit{non-trivial} bands 
cross the Fermi level. Here, we consider an extended Hubbard model on the kagome lattice,
%in the presence of spin-orbit couplings,
 involving 
a Kramers' pair of bands that have opposite Chern numbers and are isolated in the band structure. 
We construct an effective model  in a time-reversal (T) symmetric lattice description.
We determine the correlated phases of this model and identify a density-wave order in the phase diagram.
We show that this order is T-breaking, which originates from the Wannier orbitals lacking a 
common Wannier center --  a fingerprint of the underlying $Z_2$ topology.
Implications of our results for the correlation physics of the kagome metals are discussed.
\end{abstract}
\maketitle
\textit{Introduction:} 
In 
standard settings of electron correlations, the effects of short-range repulsive Coulomb 
interactions are studied on topologically trivial Bloch bands.
In the perturbative limit,  interactions are 
exactly marginal and
gapless excitations of the Fermi gas are retained in the Fermi liquid~\cite{Shankar1994}.  
In contrast, correlated phases such as Mott insulators, charge density waves
and
nematic states typically emerge 
 only when interactions become comparable to or greater than the overall 
 non-interacting electron bandwidth ($W_0$)
  \cite{Pas21.1,Ima98.1,Rozenberg1996-RMP}. 
 \par
\begin{figure}[t!]
\includegraphics[width=2.75in,height=2.25in]{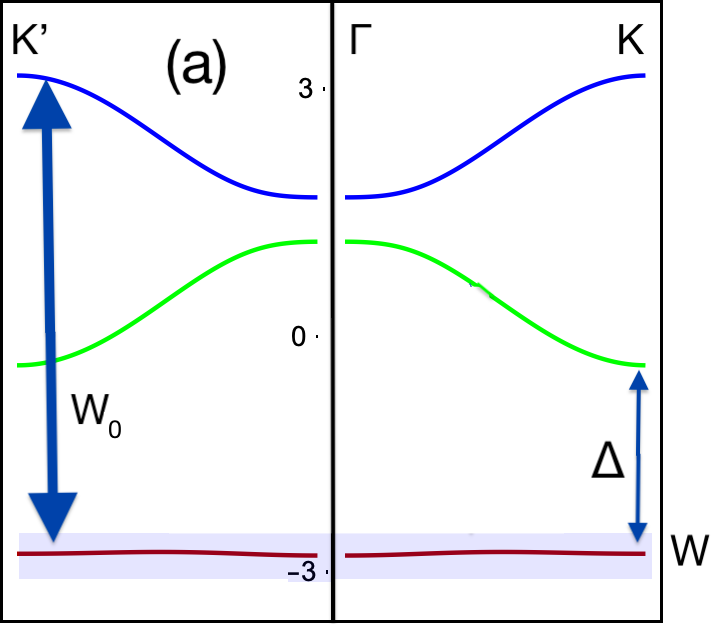}
\includegraphics[width=2.75in,height=1.5in]{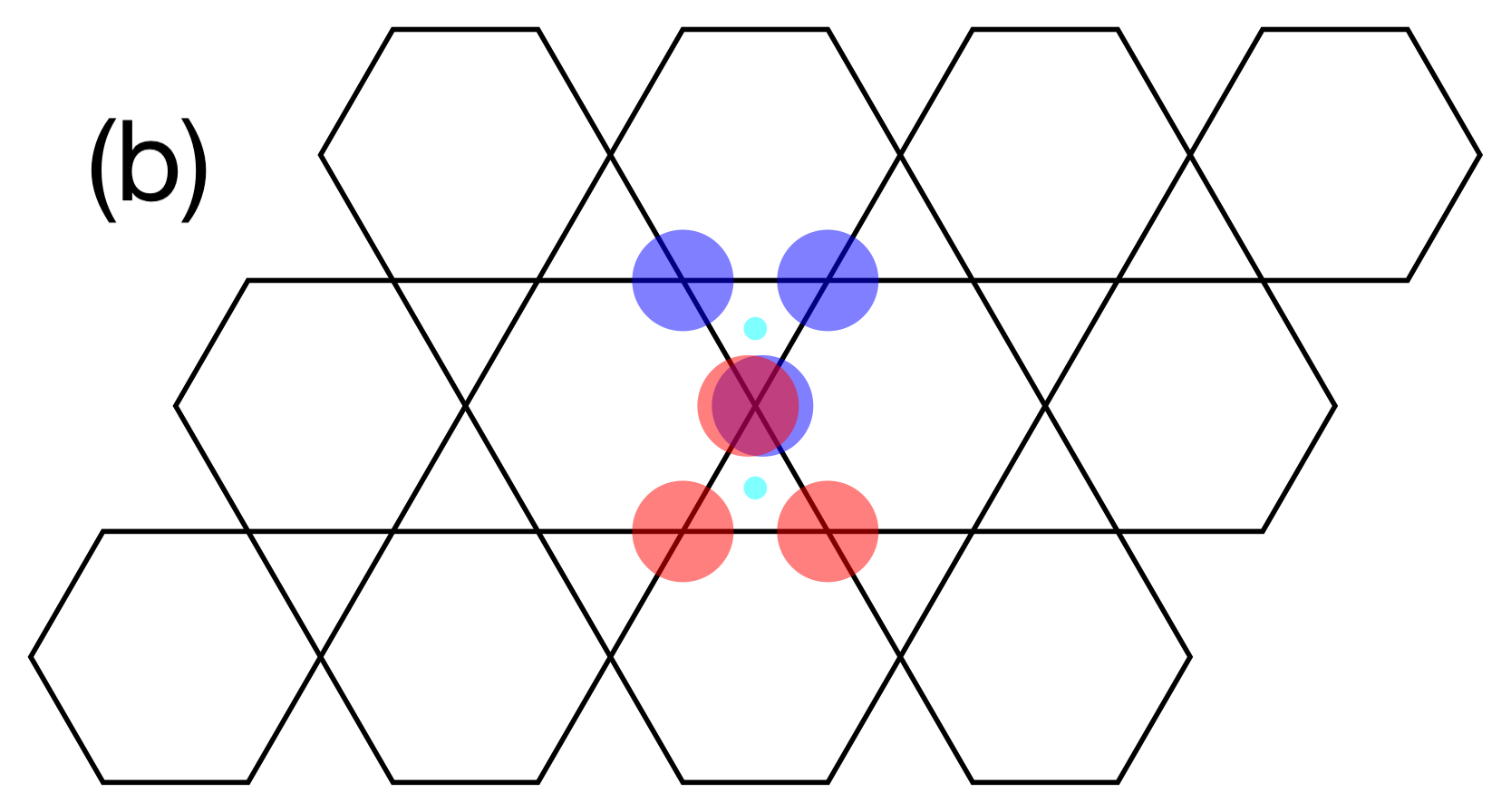} 
\caption{\textcolor{black}{(a) Plot of the non-interacting kagome bands with SOC. The effective model is projected on to the lowest pair of bands (blue rectangle) which have equal and opposite Chern numbers $C = \pm 1$.  (b) Exponentially localized Wannier orbitals (WOs) (blue and red) that no longer share the same Wannier centers (cyan dots) due to non-trivial topology.}
}\label{kagomebands}
%\vskip -0.3 cm
\end{figure}

Correlated phases
 in systems such as \textcolor{black}{twisted bilayer graphene~\cite{Pablo2018, Andrei2019} on hexagonal boron nitride (TBG/hBN)~\cite{Goldhaber2019}}, kagome lattice binary 
T$_m$X$_n$ (T= Fe, Co and X= Sn, Ge)~\cite{Checkelsky2018, Wang2020, Zhang2018, Felser2018,Yao18.1x} 
or ternary $A$V$_3$Sb$_5$ 
($A=$ Cs, K, Rb)~\cite{Hasan2020, Zeljkovic2021, Wen2021, XHChen2021, Tsirlin2021, Miao2021, Shi2021}
 intermetallics and beyond~\cite{Gil21.1x}
demand a fresh perspective to the aforementioned paradigm.
\textcolor{black}{These systems have a collection of topologically non-trivial bands
~\cite{Checkelsky2018, Comin2020, Comin2020-2, Du2018, Ortiz2020, Ortiz2021, Lei2021-Berry}
 close to the Fermi level.}
To isolate the effect of these bands, one may consider narrow bands (bandwidth $W$)
~\cite{Wang2020, Zhang2018, Sales2020, Zhang2020-PRB, Shi2021}
that are separated from the rest of the bands by a gap ($\Delta$). 
\textcolor{black}{Correlated phases
would
 already emerge when (see Fig.~\ref{kagomebands}(a))
\beq
U \sim W \ll \Delta < W_0 \quad, 
\label{Regime}
\eeq
indicating the necessity for low energy effective models obtained by projecting 
both the
 kinetic
and interacting parts of the Hamiltonian
 on to only the narrow bands. }

A key consequence
of non-trivial topology for a given band \textcolor{black}{is the well-known notion of topological obstruction ~\cite{Marzari2007, Vanderbilt2012-RMP, Vanderbilt2012} which is a double-edged sword. 
While it can substantially complicate the construction of reliable low-energy lattice
models, 
it also presents an intriguing playground to isolate and examine the novel effects of topology~\cite{Senthil2018, Senthil2019-2, Senthil2019, Randeria2021, Yang2021} on well-known phases 
of matter previously obtained purely from trivial bands. }
These issues have yet to be considered for 
kagome metals~\cite{Mielke1991, Mielke1992, Fiete10, Hu2021, Neupert2021, Nandkishore2021}, but are 
expected to be important to their understanding.  They are also general issues that are broadly important 
to the physics of electron correlations as it 
 intersects with that of electronic topology \cite{Pas21.1,Sch16.2}.

 \par   
In this Letter, we study the correlated phases of a kagome metal when its topologically nontrivial bands cross the Fermi level.
\textcolor{black}{ Importantly, we advance a non-perturbative framework
 that describes correlations in such topological metals (TMs),
especially in the regime of Eq.~\ref{Regime}. The applicability of a low energy effective lattice model in this regime with arbitrary partial fillings  distinguishes  this work from earlier attempts (see Ref.~\cite{Rachel2018} for a review) to study interactions in topological models. First, %the regime in
  Eq.~\ref{Regime} demands a non-perturbative treatment of correlations for experimentally relevant parameter values of $U, W$ and $\Delta$, which we offer here. Second, our treatment, applied in the context of partially filled Chern bands with arbitrary fillings, seeks to uncover hitherto unknown topological imprints on density wave orders. Finally, non-perturbative treatments of partially filled Chern bands that yield correlated metallic ground states (as we will see below) are exceedingly rare~\cite{Rachel2018,Fogler2002}. These features of our work stand in stark contrast to earlier attempts at deriving fractional Chern insulating phases at very specific fillings~\cite{Wen2011, Mudry2011, Sarma2011,  Bernevig2011}, or mean-field orders which are unaffected by the topology of the underlying bands in the continuum limit (for a review, see Ref.~\cite{Fogler2002}).  }\par
\textcolor{black}{To this end, we examine an extended
Hubbard model on the kagome lattice and isolate 
a Kramers' pair of bands with opposite Chern numbers crossing the Fermi level.}
Using Wannier orbitals suitable for topological bands whose nonzero Chern numbers add up to zero \cite{Vanderbilt2012},
we project the Coulomb interactions to these bands and construct an effective lattice model.
We identify a T-breaking density wave  (DW$_\text{t}$) \textcolor{black}{metal} in the phase diagram, 
 and  discuss its implications for the metallic kagome materials. \textcolor{black}{Correlated metals of this sort have not been identified from partially filled topological bands in non-perturbative settings~\cite{Rachel2018}}.  We \textcolor{black}{further}
determine how the unusual nature of this phase
is driven by 
the topological nature of 
the underlying bands and why it is absent in analogous phases derived from trivial bands.
Our
 work highlights how the obstructive power of topology can be harnessed to drive novel correlated \textcolor{black}{metallic} phases.
\par
\textit{Extended Hubbard model and topological obstruction:} 
The Hamiltonian is given by $\mathscr{H} = \mathscr{H}_0 + \mathscr{H}_I$ with
\beq \nonumber
\mathscr{H}_0 &=& -t \sum_{\langle i j \rangle \alpha \beta } c_{i \alpha }^{\dagger} c_{j \beta } +  
i \lambda_1\sum_{\langle i j\rangle } \left( \bs E_{ij} \times \bs R_{ij}\right) \cdot c_{i \alpha }^{\dagger} \bm{\sigma}  c_{j \beta } \\
&&
-t_2 \sum_{\langle \langle i j \rangle \rangle \alpha \beta } c_{i \alpha }^{\dagger} c_{j \beta } +  
i \lambda_2\sum_{\langle \langle i j \rangle \rangle  } \left( \bs E_{ij} \times \bs R_{ij}\right) \cdot c_{i \alpha }^{\dagger} \bm{\sigma}  c_{j \beta } \nonumber \\
\mathscr{H}_I &=& \sum_{i j \alpha \beta } U_{j,\alpha \beta}n_{i \alpha }n_{i+j \beta } \quad .
\label{KagomeSOC}
\eeq
The kinetic part, $\mathscr{H}_0$, describes
electrons hopping on a kagome lattice in the presence of spin-orbit couplings (SOC)~\cite{Guo09.1,Wen2011}.
Here $t(t_2)$ and $\lambda_1(\lambda_2)$ are the nearest neighbor (next-nearest neighbor) hopping 
and SOC respectively. 
$\bs R_{ij}$ is the displacement vector from site $i$ to site $j$ and $\bs E_{ij}$ is the electric field experienced along $\bs R_{ij}$. $c_{i\alpha}$ destroys an electron on site $i$ and
internal quantum number
 $\alpha$,  which refers
  to both sub-lattice and spin indices. 
The interaction part, $\mathscr{H}_I$, contains local
 interactions,
  including the on-site Hubbard interaction $U$ and the 
nearest-neighbor (next-nearest neighbor) density-density interactions $V$ ($V'$).
They are expressed in terms of $U_{j,\alpha \beta}$ that couples the number operators
$n_{i \alpha}$.
We also denote the three triangular lattice vectors $\bs n_1 = 2 a (\frac{1}{2}, \frac{\sqrt{3}}{2})$, $\bs n_2 = 2 a(-\frac{1}{2}, \frac{\sqrt{3}}{2})$, $\bs n_3 = 2 a(1,0)$ where $a$ is the nearest neighbor inter atomic spacing.
For definiteness, we
set $t_1 = 1, t_2 = -0.3,\lambda_1 = 0.28,\lambda_2=0.2$~\cite{Wen2011}.
\textcolor{black}{These parameters give rise to the non-trivial topology in the band structure (see Fig.~\ref{kagomebands}(a)) and splits the spectrum into three pairs of energetically well-separated degenerate bands with overall bandwidth $W_0\sim 6$ and inter-band gaps of order $\Delta \sim 2.33$ (between 
the middle and bottom bands).} Each pair at the top and bottom has Chern numbers $\pm 1$ while the middle pair has zero Chern number each. The spinful $\mathbb{Z}_2 $ TM
 is formed by partially filling the lowest pair of degenerate bands that have a small bandwidth $W\sim 0.04$. 
 To study the correlation effect in the regime of Eq.~\ref{Regime}, a reliable low energy effective 
 lattice
 model with exponentially decaying
 projected interactions is essential. 
Topological obstruction complicates the construction of such a model as the projection hinges on the existence of exponentially localized Wannier functions. The latter are impossible when a Bloch discontinuity exists in the BZ
  that is incurable by a unitary gauge transformation of the band(s).
  % This would the case for a single isolated topologically non-trivial band, or even a collection of bands whose Chern number sum is nonzero. 
  \par 
However, for a subspace of low-energy bands whose sum of Chern numbers vanishes, %(allowing for Chern numbers of the individual bands to be nonzero)  
there exists a unitary transformation ($U_T$) that cures all discontinuities of the relevant Bloch functions while also respecting their reciprocal lattice vector
periodicity \cite{Vanderbilt2011,Vanderbilt2012}.
\textcolor{black}{Our immediate goal, therefore, is to evaluate $U_T$ that ``smoothens" out the lowest pair of degenerate 
bands lying within an energy window $W$ such that exponentially localized Wannier orbitals can be obtained 
while also preserving the Bravais periodicity. }
%  \begin{figure}[t!]
%  %\vskip -0.3cm
%\includegraphics[width=3.5in,height=1.5in]{HaldaneRotation} \hfill
%%\includegraphics[w`idth=3in,height=3in]{u-T_PD}
%%\includegraphics[width=3.33in,height=3.33in]{PD}
%\vskip -0.9cm
%\caption{Wannier orbitals before (left) and after (right) Haldane rotation. On the left panel, 
%the Wannier orbitals share a common Wannier center and form a Kramers' pair since their corresponding 
%Bloch states satisfy Eq.~\ref{Kramers}. 
%On
% the right panel, the Wannier orbitals lack a common Wannier center and break Kramers' relations, 
% but their Bloch states instead satisfy the more general Eq.~\ref{TRS}. 
%}\label{HaldaneRotation}
%\vskip -0.3 cm
%\end{figure}
\textcolor{black}{The procedure for evaluating $U_T$ %and thereby obtaining exponentially localized Wannier functions 
using the parallel transport gauge 
is briefly summarized in the Supplemental Material (SM) Sec.~A. }
%The basic idea involves constructing two Bloch states, $|u_1\rangle$, $|u_2\rangle$ with Chern numbers $\pm 1$, 
%smooth on a cylinder, and then using $U_T \equiv U_H$ to rotate them into final states, $|u_{f1}\rangle$, $|u_{f2}\rangle$, 
%that are trivial and smooth on a torus. 
%\textcolor{black}{Here, $U_H$ is defined as the matrix of eigenstates of the Haldane model also properly smoothened on a cylinder. } 
%Some of the real and imaginary components of  $|u_{f1}\rangle, |u_{f2} \rangle$ are plotted in 
%Fig.~\ref{Bloch} in SM. 
%%Fig.~3 in the SM. 
 \par
\begin{figure}[h!]
  %\vskip -0.3cm
\includegraphics[width=3.5in,height=1.5in]{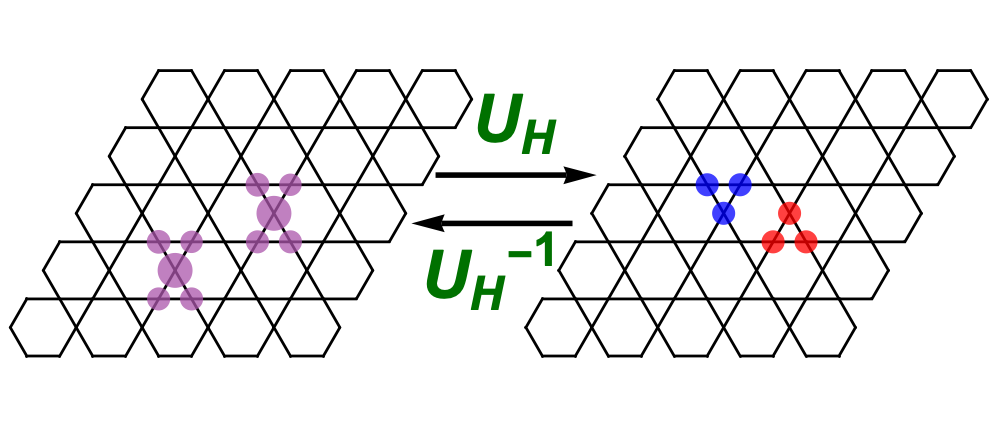} \hfill
\vskip -0.9cm
\caption{Wannier orbitals before (left) and after (right) Haldane rotation. On the left panel, 
the Wannier orbitals share a common Wannier center and form a Kramers' pair. 
On
 the right panel, the Wannier orbitals lack a common Wannier center and break Kramers' relations, 
 but their Bloch states instead satisfy the more general Eq.~\ref{TRS}. 
}\label{HaldaneRotation}
\vskip -0.3 cm
\end{figure}

 
 \textcolor{black}{
\textit{Wannier orbitals:} 
We denote the final smoothened Bloch states (Figs.~\ref{Bloch},~\ref{Decay},~\ref{fig:loc_wo} of SM) as $|u_{f1}\rangle, |u_{f2} \rangle$,  obtained from the initial states $|u_1\rangle$, $|u_2\rangle$ with Chern numbers $\pm 1$. The transformation that achieves this is  $U_T \equiv U_H$, where $U_H$ is defined as the matrix of eigenstates of the Haldane model also properly smoothened on a cylinder. }
%Fig.~3 in the SM. 
%There are several notable features about these Bloch states.
%As the procedure involves a series of unitary transformations, the states remain degenerate and orthonormal. 
%Given their individual Chern numbers are zero and the states are smooth on the torus,
 %their Wannier functions are exponentially localized in real space 
 %(See Figs.~\ref{fig:loc_wo}, ~\ref{Decay} and \ref{Bloch} of SM). 
 %(See Figs.~6, ~4 and 3 of SM). 
 %Since the original states 
 %$|u_{1}\rangle, |u_{2} \rangle$ are eigenstates of the original Hamiltonian, 
 %so are $|u_{f1}\rangle, |u_{f2} \rangle$ which reproduce the lowest two bands of the kagome band structure. 
 %This 
%would not be
 %true if $|u_{1}\rangle, |u_{2} \rangle$ 
%were
  %not degenerate. 
 In Fig.~\ref{HaldaneRotation} of SM
   the geometric effect of applying $U_H$ 
  on the Wannier orbitals of $|u_{1}\rangle, |u_{2} \rangle$ is depicted.  
  Before the action of  $U_H$, the two Wannier orbitals share the same Wannier center, 
  have equal onsite energies, and the corresponding Bloch states satisfy the Kramers' relations, $T |u_{1} \rangle_{\bs k}  = |u_{2}\rangle_{-\bs k}$ and $T |u_{2} \rangle_{\bs k} = -|u_{1}\rangle_{-\bs k}$,
%\beq \nonumber
% T |u_{1} \rangle_{\bs k}  &=& |u_{2}\rangle_{-\bs k}\\
% T |u_{2} \rangle_{\bs k}  &=& -|u_{1}\rangle_{-\bs k} \quad , 
% \label{Kramers}
% \eeq
where $T$ is the time reversal operator. After the application of $U_H$, $|u_{f1}\rangle, |u_{f2} \rangle$ 
no longer satisfy the relations and hence do not form a Kramers' pair. 
\textcolor{black}{The two corresponding Wannier orbitals have different Wannier centers, as shown in SM Fig.~\ref{HaldaneRotation} (right panel) or Fig.~\ref{kagomebands}(b), 
but are smooth in the momentum space and exponentially localized in the real space. }

Denoting the operators 
for the Wannier orbitals in the new basis as $b_{\bs k \mu}$ ($\mu=1,2$), the following transformation properties under time-reversal hold
\beq
T \hat{\Psi}_{\bs k} T^{-1} = \hat{\mathscr U}(\bs k) \hat{\Psi}_{-\bs k} \quad .
\label{TRS}
\eeq
Here $\hat{\Psi}_{\bs k}$ is a column vector of the operators $b_{\bs k \mu}$ and $\hat{\mathscr U}(\bs k)$ 
is a unitary matrix formed from sums of products of $U_H$ matrix elements (see SM Sec.~A). 
Due to the aforementioned properties, the set of Bloch states $|u_{f1}\rangle, |u_{f2} \rangle$ 
now forms a convenient basis~\cite{FootNote} to write an effective low-energy model with a minimal set of interaction and 
hopping parameters that respect translations.  
%Attempting to write an effective model from the original states $|u_{1}\rangle, |u_{2} \rangle$ is problematic 
%as the corresponding hopping and interactions are not exponentially decaying, and would be 
%unsuitable to describe correlated states~\cite{FootNote}.
We also note that despite $|u_{f1}\rangle, |u_{f2} \rangle$ having zero Chern numbers, 
the topological imprints of the original $\mathbb{Z}_2$ TM are now encoded in the lack of a common Wannier center 
between the Wannier orbitals and revised definition of the time-reversal operator Eq.~\ref{TRS}. 
Demanding time-reversal symmetry in the effective two-orbital Hamiltonian leads to additional non-local terms driven purely by the topology of the underlying bands and absent for trivial bands. 
 \par  
 \textit{Projected lattice Hamiltonian:} We are now equipped to project the
local interactions  $U$, $V$ and $V'$ of the original model, Eq.\,\ref{KagomeSOC},
onto the bands $|u_{f1}\rangle, |u_{f2} \rangle$. We denote $b_{i\mu}$ 
as the operator that destroys an electron at site $i$ and Wannier orbital $\mu = 1,2$ (WO 1, WO 2 for brevity).
The total effective two orbital Hamiltonian ($H$) after projection is written as a sum of non-interacting band term ($H_0$) and interactions ($H_I$) given by 
\beq \label{Hamiltonian}
H &=& H_0 + H_I  
\label{eq:hamiltonian_eff}
\\
H_0 &=& 
%\sum_{\substack{ij \\ \mu \nu}} 
\sum_{ij} \sum_{\mu \nu}
b_{i \mu}^\dag t_{ij}^{\mu \nu} b_{j \nu} 
\nonumber
 \\ 
 H_{I} &=& 
 %\sum_{\substack{ijkl\\\mu\mu'\nu\nu'}}
 \sum_{ijkl} \sum_{mu\mu'\nu\nu'}
  u_{\mu\mu'\nu\nu'}(j,k,l) b_{i\mu}^\dag b_{i+j\mu'}
 b_{i+k\nu}^\dag b_{i+l\nu'}  \quad . \nonumber
%H_0 &=& \sum_{\bs k \mu \mu'} \epsilon_{\mu \mu'} (\bs k)  d_{\bs k \mu}^{\dagger} d_{\bs k \mu'} \\ \nonumber
%H_I &=& \sum_{\substack{\mu \nu \\ \mu \neq \nu}}\sum_{\bs k \bs p \bs q} U'^{\mu \nu \nu \mu} d^{\dagger}_{\bs p -\bs q\mu}
\eeq
Here $t_{ij}^{\mu \nu}$ and $u_{\mu\mu'\nu\nu'}(j,k,l)$ are respectively 
the hopping parameters and interaction matrix elements, which
%The various hopping and interaction parameters 
are given in the SM (Tables I-III). 

Several points are in order.
First, in writing the interaction terms, transformation properties of the individual WOs under rotations and reflections must be considered in addition to the Bravais lattice symmetries. These properties are especially non-trivial since the two WOs no longer share the same Wannier center.  
All symmetries of the kagome lattice are respected once the transformation properties 
of individual WOs are taken into account. Second, 
although the two WOs now no longer form a Kramers' pair, 
the interaction Hamiltonian preserves time-reversal symmetry. The non-trivial effect of topology is now manifest in the 
non-local terms that follow from time-reversal invariance as well as the 
lack of a common center for the two WOs. 
With respect to the hopping parameters, the low-energy band structure can already be reproduced 
(see SM Fig.~\ref{BSComparison}) 
%(see SM Fig.~5) 
from the effective model by truncating hopping parameters 
of the two WOs to the third neighbors.
\par  
 \begin{figure}[t!]
\includegraphics[width=3.3in,height=2.2in]{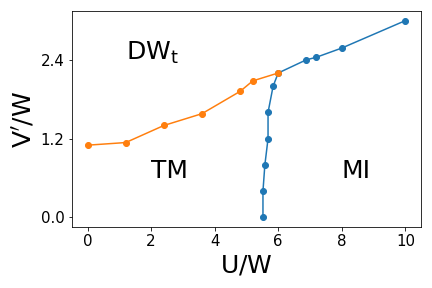} \hfill
\vskip -0.3 cm
\caption{
Phase diagram of the model Hamiltonian in Eq.~\ref{Hamiltonian} 
expressed in terms of the interaction parameters of the original model, $U$ and $V'$,
in the unit of bandwidth $W$ of the projected bands.
The blue 
%line 
and orange
\qs{ lines}
 denote first-order
 % transition
  and second-order
  \qs{ transitions}
  respectively. 
  %Three phases in the
  \qs{The}
   phase diagram
   % are
   \qs{involves}
    topological metal (TM), $T$-breaking
density wave order (DW$_\text{t}$),  and Mott insulator (MI). 
In terms of the effective interactions of Eq.~\ref{eq:hamiltonian_eff} (see Tables II and III, SM),
the threshold values
are $\sim 0.35$ ($\sim 0.14$) along the horizontal  (vertical) axis.
 }\label{PD}
 \vskip -0.3 cm
\end{figure}
\textit{Phase diagram and the nature of the DW$_{\text{t}}$ order:}
We investigate the correlation effects via a $U(1)$ slave-spin method~\cite{SlaveSpin}
 (see SM, Sec.~E).
%In this approach, we rewrite the electron operators as $b_{i\mu} = S_{i\mu}^- f_{i\mu}$ 
%and solve the model at saddle point level. 
\textcolor{black}{In the $U(1)$ slave spin method, we rewrite the electron operators as
$b_{i\mu} =S^{-}_{i\mu} f_{i\mu }$
 where $f$ is the fermion operator and $S^-$ is the spin operator.  The original Hilbert space can be mapped onto an extended space of slave spins and $f$ fermions according to the correspondence:
$|1\rangle_b = |1\rangle_f |\uparrow\rangle_{S}$
$|0\rangle_b = |0\rangle_f |\downarrow\rangle_{S}$. 
To project out the unphysical degree of freedom, we enforce the constraint:
 $S^{z}_{i\mu}+\frac{1}{2} =f_{i\mu}^\dag f_{i\mu }$.
The slave spins carry the $U(1)$ charge which allows us to rewrite the four-WO interaction terms in Eq.~\ref{eq:hamiltonian_eff} as  
%\begin{eqnarray} 
$b_{i\mu}^\dag b_{i\mu} b_{j\mu'}b_{j\mu'} = (S_{i\mu}^z +\frac{1}{2})(S_{j\mu'}^z +\frac{1}{2}).$
%\end{eqnarray} 
We then perform a 
%mean field 
\qs{Hubbard-Stratonovich}
decoupling of the spin terms above and solve the model at 
\qs{the}
saddle point level. The full Hamiltonian written in the slave-spin basis along with the self-consistent parameters is given in Sec.~E of SM .
 We note that, due to the offset in the Wannier centers ({\it c.f.} Eq.~\ref{eq:approx_int}, SM), the
 nearest neighbor interaction
 $V$ plays a similar role as $U$ in the effective model
and, in particular, its tendency to drive a density wave order with an ordering wave vector $K$ is reduced.}
 Instead, $V$  only modifies the phase boundaries, 
 so we have set  it to zero  for convenience. 

The phase diagram at half-filling is shown in Fig.~\ref{PD}. 
The horizontal and vertical axes are marked by the onsite and next-nearest neighbor interactions, 
$U$ and $V'$ respectively. 
 The phase diagram contains three phases: a TM,  Mott insulator (MI), and 
 a density wave phase (DW$_{\text{t}}$),
which has an ordering wave vector $M$.  There are also three phase boundaries,
 and a
  tri-critical 
  phase coexistence
  occurs around $(U/W,V'/W) \sim (6, 2.3)$. 
 In the TM and DW$_{\text{t}}$
 phases, the
  quasi-particle weights remain
 nonzero and the system is metallic.
 In the 
 MI phase, 
 the
 quasi-particle weights vanish due to the interactions, and the electrons are localized.  
 The Mott phase is also
 spontaneously T-breaking and characterized by a fully 
 polarized ferro-orbital order in the WO basis which yields an in-plane spin order in the original kagome basis 
 of Eq.~\ref{KagomeSOC}. 
\par 
We now turn to
discussing
 the nature of the DW$_{\text{t}}$ phase.  \textcolor{black}{ This phase 
is a charge order when viewed in the WO basis with a modulation vector at the $M$ point. This is interesting since, in the absence of topology, a Kramers pair of WOs share the \textit{same} Wannier center and a minimization of the onsite Coulomb repulsion would have instead yielded a 
\qs{(pure)}
spin order. In the current situation, due to the lack of a common center from the original topological bands, the  \textit{non-local} $V'$ interaction instead drives a charge order at $M$ point. Now when}
 \textcolor{black}{viewed in the original
   kagome basis,  the DW$_t$ contains a ``collinear charge order" -- a phase where the onsite charge density is modulated along the M direction; i.e., it acquires a gradient along one of the crystal axis direction but not the other -- that is entwined
with a collinear in-plane spin order
 by the $\mathbb{Z}_2$ topology. Due to the coexistence of the collinear charge and spin orders,  the DW$_{\text{t}}$  phase  is naturally T-breaking.}
Such a property is absent in correlated phases derived from trivial bands. 
To see this, we write the charge operator and spin operator along $x$ direction
 in the kagome basis as $\phi_{\bs q}^c = \sum_{\alpha} c^\dag_{\bs k\alpha} 
 c_{\bs k+ \bs q \alpha}$ and $\phi_{\bs q}^x = \sum_{\alpha\gamma} \sigma^x_{\alpha\gamma}
 c^\dag_{\bs k\alpha} c_{\bs k+ \bs q \gamma}$ respectively. 
 Switching to WO basis we obtain (see Sec.~D of SM) 
 \textcolor{black}{
\beq \nonumber
\phi_q^c& \sim &  
 F(\bs n_1, \bs n_2) A^2(n_{1,q} +n_{2,q} ) 
 +
G(\bs n_1, \bs n_2) A^2(n_{1,q} -n_{2,q} ) \quad , \nonumber \\
 \phi_q^x &\sim &
G(\bs n_1, \bs n_2) A^2(n_{1,q} +n_{2,q} ) \nonumber 
 +
F(\bs n_1, \bs n_2) A^2(n_{1,q} -n_{2,q} )  \quad .
 \label{eq:dw-mixing}
\eeq
where $A \sim 0.5$ is the overlap constant, $n_{\bs q j}$ is the density operator for WO $j$ and $F(\bs n_1, \bs n_2) = \frac{4+e^{iq\bs{n}_1}+e^{-iq\bs{n}_2} }{2}$, $G(\bs n_1, \bs n_2) =  \frac{2-e^{iq\bs{n}_1}-e^{-iq\bs{n}_2} }{2} $.}
 Consider the DW$_{\text{t}}$ appearing in the phase diagram Fig.~\ref{PD}. This phase is a charge order in the WO basis with 
 $n_{\bs{q}1}+n_{\bs{q}2} \sim \eta \delta_{\bs q,M} $ and $n_{\bs{q}1}-n_{\bs{q}2}\sim 0 $. 
 Switching to the original kagome basis,
 it contributes to both charge and spin channels.
 A real space illustration of the intertwined nature of charge and spin degrees of freedom 
 (defined with respect to the original kagome operators) is shown in
  Fig.~\ref{fig:real_space_pattern} of SM.
 % Fig.~7 of SM.
 The $\bs{n}_1$ and $-\bs{n}_2$ 
 vectors 
 in the phase factors of Eq.~\ref{eq:dw-mixing} 
  reflect the difference in the Wannier centers of the two WOs (see SM, Sec.~C).
 \textcolor{black}{Had they vanished, as for the WOs of topologically trivial bands, the mixing of the spin order into the charge order 
 would have been absent, \textit{despite} a non-zero SOC~\cite{FootNote2}. }
Hence this property is a direct consequence of the non-trivial $\mathbb{Z}_2$ topology of the kagome Hamiltonian. \par %\textcolor{black}{We further stress that  SOC is not directly responsible for the coupling of spin and charge degrees.   First, we clarify that, in our case, the order parameter for the charge order is a scalar. Hence it does not couple to the spin order through SOC. This is unlike the vectorial orbital loop current order parameters studied in, for example, Klug et al Phys. Rev. B 97, 155130 (2018).  Finally, the fact that the WO charge order implies both spin and charge order in the original basis in our model is not a \textit{direct} consequence of SOC. Instead, it is a direct consequence of the topological non-triviality of isolated bands. Such intertwined order would have been absent if the bands were topologically trivial or overlapping even in the presence of SOC.} \par 
    \textit{Discussions:}
   Several remarks are in order. First, the  DW$_\text{t}$
order is collinear and has an
   ordering wave vector $M$. \textcolor{black}{At this level of energetics, all three $M$ point orders in the BZ are feasible and hence degenerate.  Further breaking of this degeneracy can occur once perturbations from coupling to the lattice via structural distortions is accounted for.} 
   \qs{In particular, such subleading energetics can lock the system into a 3$Q$ order, which has 
   the $2 \times 2$ real space pattern as observed in AV$_3$Sb$_5$ \cite{Hasan2020, Zeljkovic2021}.}
   \textcolor{black}{We can additionally expect such an order to form 
   a composite field that serves as an accompanying $C_3$-breaking
   nematic order parameter, as has been found in recent experiments in KV$_3$Sb$_5$~\cite{Zeljkovic2021-Nematic} and studied in the context 
   of the moire systems \cite{Che20.1x}}. We reserve a systematic study of the nematicity for a \textcolor{black}{later} work.
   \par
   Second, we have considered a regime of parameters which facilitates a controlled theoretical analysis, 
   with a particular form for SOC.
   Because the DW$_\text{t}$
order appears as a phase, it is expected to be robust over an extended 
    range
   of parameters and SOC types.
    Thus, even though in $3d$-electron systems
   such as
   T$_m$X$_n$ (T= Fe, Co and X= Sn, Ge)
   and
    $A$V$_3$Sb$_5$ ($A=$ Cs, K, Rb),
   the SOC is typically several percents of the overall bandwidth, we expect 
   DW$_\text{t}$
    to be relevant.
  Finally, due to the involvement of Haldane eigenstates in the construction of the Hamiltonian Eq.~\ref{Hamiltonian}, 
 for ground states that feature non-trivial boundary physics 
 one must treat the boundary physics by reverting to 
  the original kagome basis.
  \par
   \textit{Implications for Experiments:}
     Several correlated phases~\cite{Hasan2020, Zeljkovic2021, Wen2021, XHChen2021, Tsirlin2021, Miao2021, Shi2021}
	have been observed in 
	$A$V$_3$Sb$_5$.
	%($A$VS, $A=$Cs, K, Rb).
     These systems have topological bands near  their Fermi level, a key ingredient for the relevance 
     of the 
     DW$_\text{t}$ phase.
     However, 
     \qs{the width of}
     these topological bands 
     %are wide
     \qs{is intermediate} and the strength of the correlations 
     appears to be relatively weak, which suggests that the mixing of the spin order in the underlying charge density wave order 
     will be relatively weak.
     These features are qualitatively consistent with the experimental observations.
     The charge order is indeed centered at the $M$ point of the BZ~\cite{Hasan2020, Zeljkovic2021, Wen2021, Miao2021}. \textcolor{black}{Both
   scanning tunneling microscopy (STM)~\cite{Hasan2020} and anomalous Hall transport~\cite{Ali2020-2} measurements find evidence of chiral topological charge order
    that breaks time-reversal symmetry}.  In addition, $\mu$SR has so far found no evidence of static moments~\cite{Graf2021}. 
    Thus, any spin order
   must be below the $\mu$SR resolution limit
   \qs{(see, however, ``Note added" below).}
%\par
In the kagome lattice binary T$_m$X$_n$ (T= Fe, Co and X= Sn, Ge) 
compounds~\cite{Checkelsky2018, Wang2020, Zhang2018, Felser2018}, 
flat bands are located about $\lesssim 0.3$ eV from 
the Fermi level~\cite{Comin2020, Comin2020-2, Wang2020}, and
the topological nature of the Fermi-level-crossing bands remains to be established. 
Because the correlation strength here is expected to be considerably larger than the $0.3$ eV flat-band energy scale,
we expect the latter to significantly influence the mechanism and nature of the correlated phases. Our work motivates
the search for T-breaking charge-density-wave phases in the T$_m$X$_n$ and related systems.
\par
   \textit{Conclusions:}
   We studied correlation effects of a kagome metal when a pair of $Z_2$ topological bands with nonzero but 
    opposite Chern numbers cross the Fermi level. 
    Using a time-reversal symmetric lattice description, we constructed an effective lattice model and 
    determined its correlated phase diagram. 
    We identified a density-wave order that breaks time-reversal symmetry (DW$_\text{t}$)
 in the phase diagram. 
    The phase originates from a Coulomb avoidance in a way that respects the topology of the underlying bands. 
    We expect this mechanism to operate in broader contexts of correlated topological systems. 
    The T-breaking density-wave order has important implications for understanding rapidly emerging 
experiments in the ternary kagome metals. Such phases may also be explored by future experiments in 
the binary kagome systems as well as in other correlated metals whose crystalline lattice nurtures topological bands.\par

%\acknowledgements
\begin{acknowledgments}
\textit{Acknowledgements:} We thank P. C. Dai,
\qs{Z. Guguchia}
and M. Yi for useful discussions. This work has in part been supported by
the U.S. Department of Energy, Office of Science, Basic Energy Sciences, under Award No. DE-SC0018197
and  the Robert A.\ Welch Foundation Grant No.\ C-1411 (Q.S.).
Q.S. acknowledges the hospitality of the Aspen Center for Physics,
which is supported by NSF grant No. PHY-1607611.
\end{acknowledgments}
\textcolor{black}{\textit{Note added}: After
% the completion of
 this manuscript
 \qs{became available on the arXiv,}
  two new $\mu$SR~\cite{Guguchia2021-TRSB, Zhao2021-TRSB} and polar Kerr~\cite{Wang2021-Kerr} measurements found evidence of broken time-reversal symmetry in the density wave phase of ternary Kagome systems. The $\mu$SR data detect both in-plane and out-of-plane components of the broken time reversal symmetry supporting our findings. }

\noindent
$\oplus$ These authors contributed equally to this work.\\
$\dagger$ csetty@rice.edu

\par
\bibliography{Kagome.bib} 

\begin{thebibliography}{61}
\bibitem [{Foo({\natexlab{a}})}]{FootNote}%
  \BibitemOpen
  \href@noop {} {\bibfield  {journal} {\bibinfo  {journal} {It is worth
  emphasizing that, since $U_H$ is a unitary gauge transformation that is
  smooth only on a cylinder due to the non-trivial topology of the Haldane
  model, observables of two Hamiltonians formed by the two sets of Wannier
  orbitals are only invariant in the absence of a boundary. Hence the
  predictions of the two models -- provided one is able to construct a low
  energy Hamiltonian from $|u_{1}\rangle, |u_{2} \rangle$ -- are the same as
  long as only bulk properties are probed}\ } ({\natexlab{a}})}\BibitemShut
  {NoStop}%
\bibitem [{Foo({\natexlab{b}})}]{FootNote2}%
  \BibitemOpen
  \href@noop {} {\bibfield  {journal} {\bibinfo  {journal} {Since the order
  parameter for our charge order is a scalar, it does not couple to spin
  degrees through SOC. This is unlike other vectorial density-wave order
  parameters studied in the context of loop current orders.}\ }
  ({\natexlab{b}})}\BibitemShut {NoStop}%
\end{thebibliography}
\newpage
\onecolumngrid
\newpage
%\begin{multicols*}{1}
\section{Supplemental Material }
\subsection{A. Parallel transport}
\begin{figure}[b!]
\includegraphics[width=3in,height=2in]{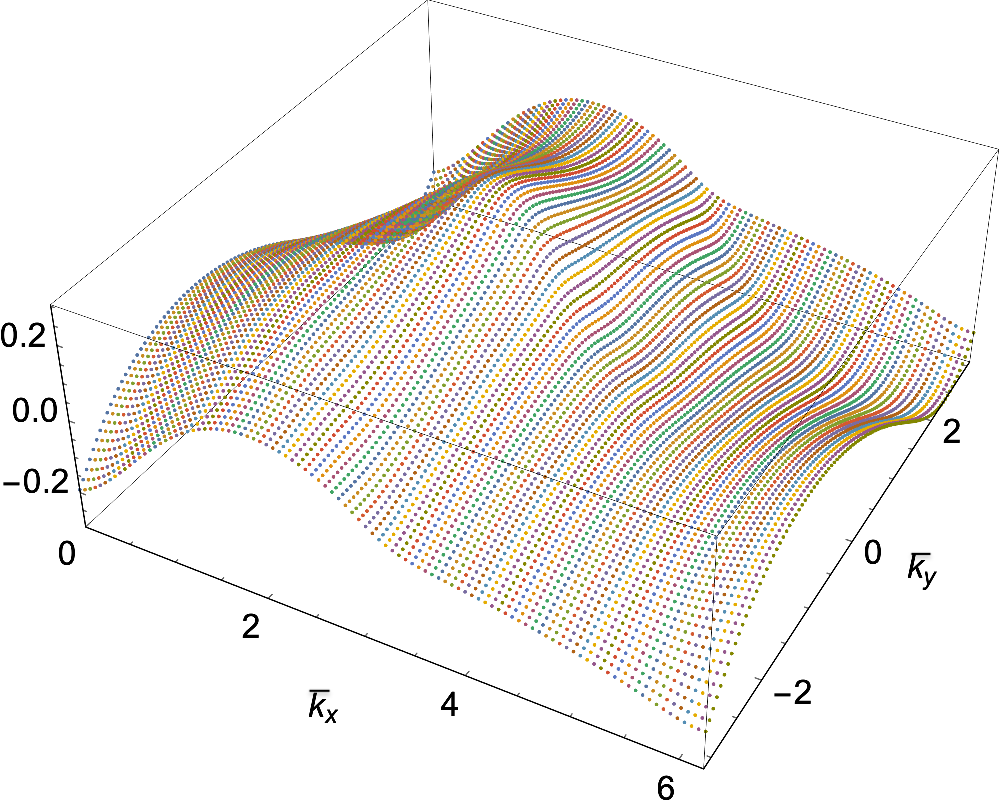}
\includegraphics[width=3in,height=2in]{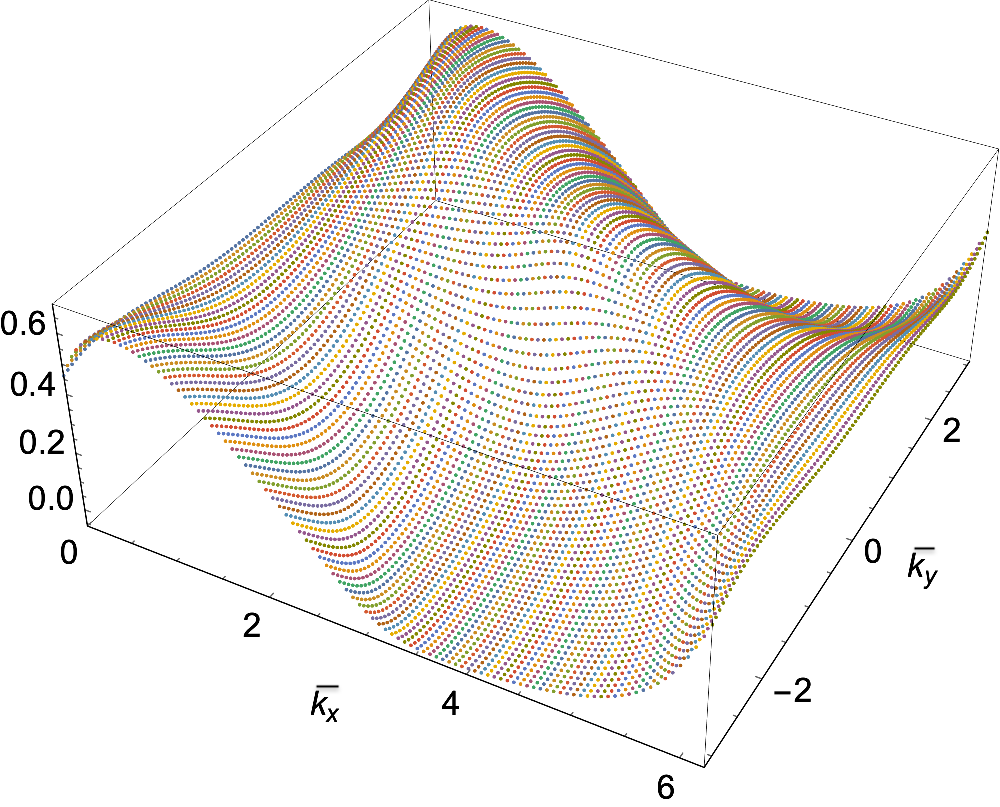} 
\includegraphics[width=3in,height=2in]{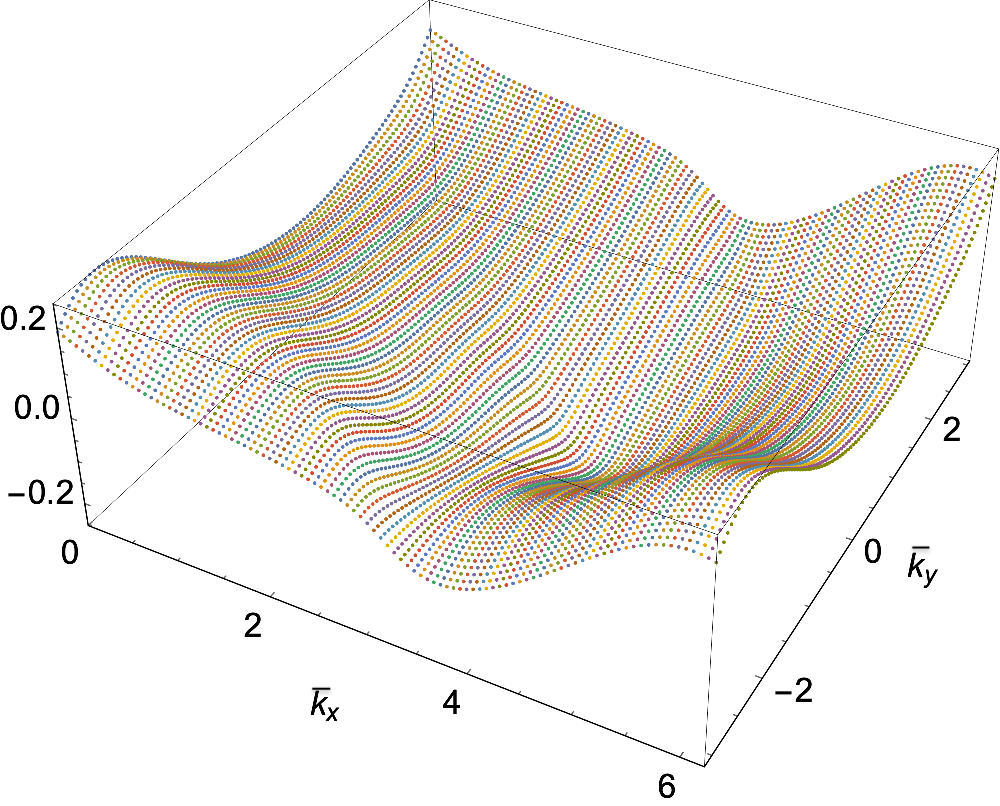} 
\includegraphics[width=3in,height=2in]{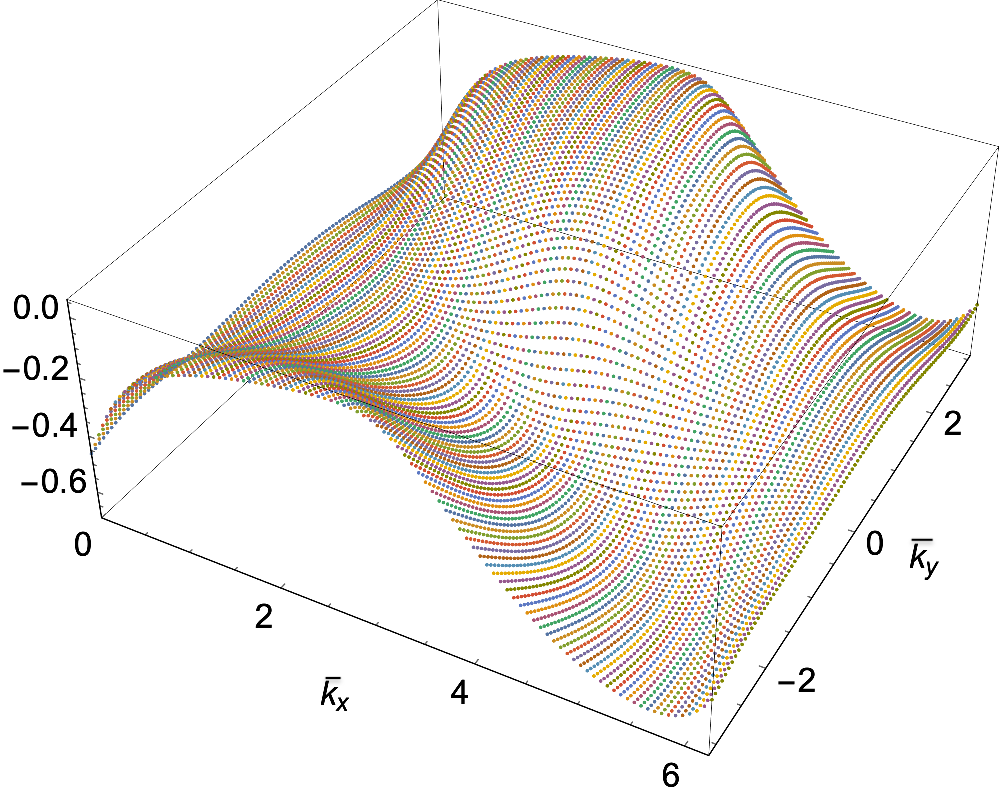} 
\caption{Example plots of Bloch functions in the rescaled and rotated rectangular Brillouin zone after the Wannierization process. (Top row) Real (left) and imaginary (right) parts of fourth component of $|u_{f1}\rangle$. (Bottom row) Same as top row but for $|u_{f2}\rangle$.}  \label{Bloch}
\end{figure}

\begin{figure}[h!]
\includegraphics[width=3.5in,height=2.25in]{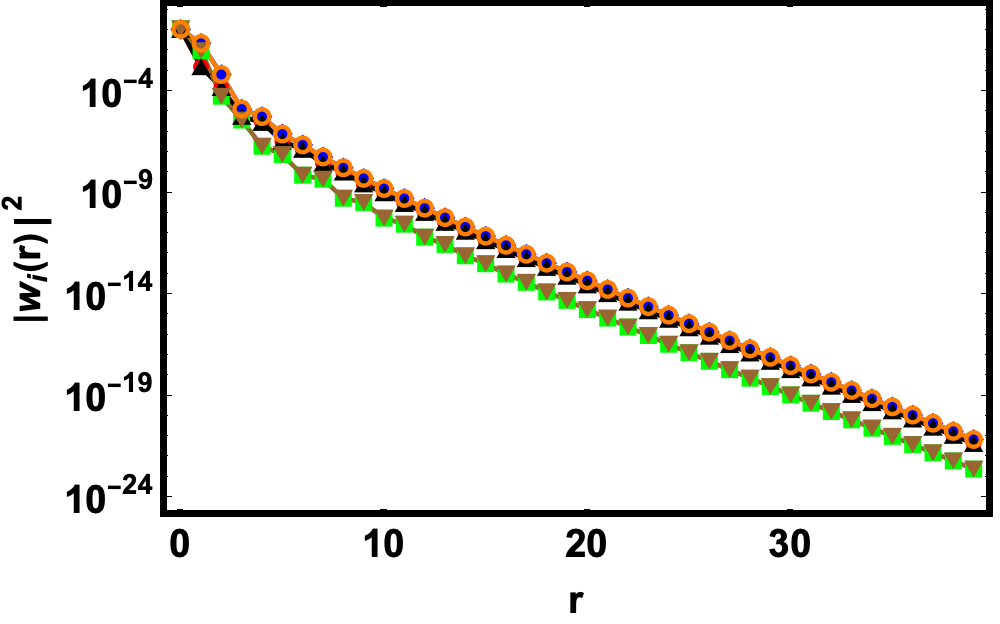}\hfill
\includegraphics[width=3.5in,height=2.25in]{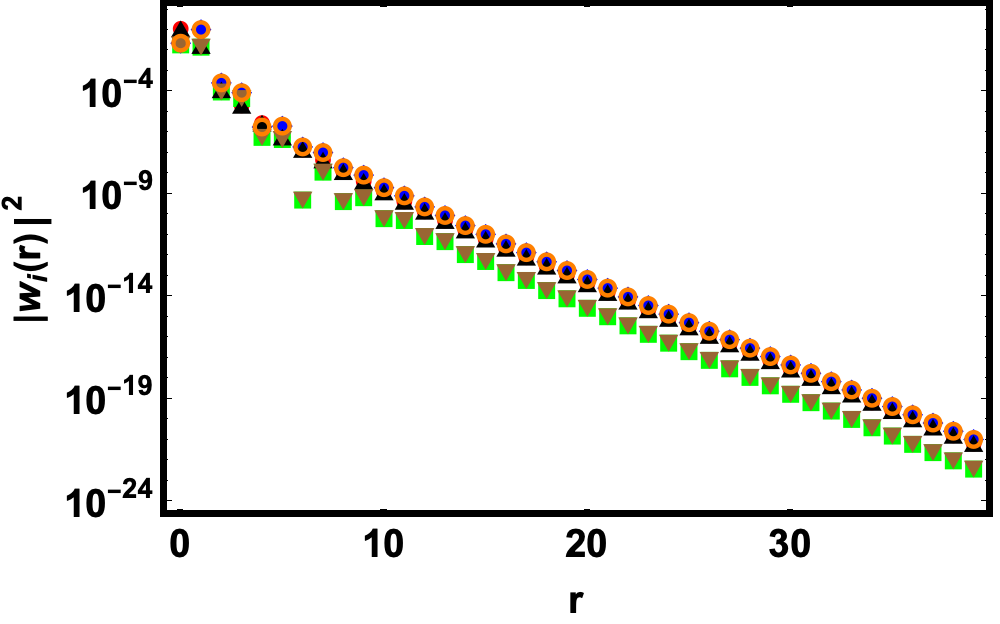}\hfill
\caption{Plots of probability density along the six projections of the first (left) and second (right) Bloch states as a function of distance from Wannier center. }   \label{Decay}
\end{figure}

Topological obstruction is the lack of a smooth Bloch representation for a given band throughout the Brillouin zone (BZ) which also respects symmetries of the Bravais lattice. Such obstruction, depending on the type,  can substantially complicate the construction of reliable low-energy lattice models that are necessary to treat the correlated system's need to avoid the Coulomb repulsion. In situations which the obstruction can be resolved, like the specific case we discuss in this work, a key task is to determine the unitary band transformation matrix $U_T$ for each $\bs k$ point in the Brillouin zone grid that untangles the topology. Below we outline the steps for evaluating the form of $U_T$ and applying it to the Bloch states in our problem.   \\ \newline
Determining the form of $U_T$ 
for specific systems can be a non-trivial task that depends on the lattice structure and the degree of Wannier localization desired. 
The simplest procedure~\cite{Vanderbilt2011} involves projecting the $n$ Bloch states onto an \textit{ansatz} 
of $n$ trial states and using the determinant of the resulting $n \times n$ matrix to renormalize the original Bloch functions. 
This procedure was used to derive reliable tight-binding description 
in moire
 superlattices~\cite{Senthil2018, Senthil2019-2, Senthil2019} and 
 recently implemented in Lieb lattices~\cite{Randeria2021} 
to derive bounds on superfluid stiffness.  An obvious difficulty with the projection method, however, 
is the requirement of a suitable \textit{ansatz} with the correct symmetries. The initial guess becomes harder with more quantum numbers, increased size of Hilbert space and complex lattice geometry. In the case of the kagome lattice, even if the correct symmetry of the wave functions is used as a guess,  non-singular wave function normalization throughout the BZ is not assured. To circumvent this difficulty, we instead use the more intuitive parallel transport gauge~\cite{Vanderbilt2012}
 from which $U_T$ can be obtained more readily.   \\ \newline
 In this section, we briefly outline the various steps involved in obtaining the parallel transport gauge used for the kagome problem and refer the reader to Ref.~\cite{Vanderbilt2012} for further details.  First, through a linear transformation, the two low-lying Bloch functions, $|u_1\rangle, |u_2 \rangle$, with Chern numbers $\pm 1$ and bandwidth $W$ are redefined on a rectangular BZ where $\tilde k_x \in [0,2\pi]$, $\tilde k_y \in [-\pi, \pi]$.  Next, the states are parallel transported along $\tilde k_y =0$ and the continuity between $\tilde k_x = 0$ and $\tilde k_x = 2 \pi$ is restored by redistributing the total phase along the  $\tilde k_y =0$  line. In the third step, for every $\tilde k_x$, two directionally opposite parallel transports are carried out -- one from $(\tilde k_x, 0)$ to $(\tilde k_x, \pi)$, and another from $(\tilde k_x, 0)$ to $(\tilde k_x, -\pi)$.  In the next step, periodicity along $\tilde k_y$ is restored at $\tilde k_x =0$ and $\tilde k_x = 2\pi$ while an off-diagonal unitary matrix, $V(\tilde k_x)$, connects the two states for every remaining $\tilde k_x$. In the penultimate step, the two Bloch states are untangled by diagonalizing the matrix $V(\tilde k_x)$ without disturbing their smoothness in rest of the BZ. At this stage, one is left with a Kramers'
  pair of bands with Chern numbers $\pm 1$ that is periodic with respect to reciprocal vectors along $\tilde k_x$ but not along $\tilde k_y$ for each $\tilde k_x$ except $\tilde k_x =0, 2\pi$, i.e., they are smooth on a cylinder.  In the final step, the Bloch states are rotated into a gauge that is smooth on a torus by using the eigenstates of the Haldane model ($U_H$) that have Chern numbers $\pm 1$. The Haldane matrix $U_H$, properly smoothened on a cylinder, plays the role of $U_T$ and rotates the original Bloch states into a trivial basis  $|u_{f1}\rangle, |u_{f2} \rangle$ with zero Chern numbers each. The topological obstruction is thus resolved. Certain real and imaginary components of $|u_{f1}\rangle$ and $|u_{f2}\rangle$ are plotted in Fig.~\ref{Bloch} and the exponential decay of the Wannier functions is shown in Fig.~\ref{Decay}. \\ \newline
The linear transformation matrix $U_T$ is defined as the matrix ($U_H$) of eigenvectors of the Haldane model properly smoothened on a cylinder according to the prescription detailed above. We denote these matrix elements as $g_{ij}(\bs k)$. Using this matrix for $U_T$
\begin{gather}
U_T \equiv U_H
 =
  \begin{bmatrix}
  g_{11} (\bs k)&
g_{12} (\bs k)\\
 g_{21} (\bs k)&
 g_{22} (\bs k)
   \end{bmatrix},
\end{gather}
the unitary matrix $\mathscr{\hat{U}} (\bs k)$ which defines the action of time reversal operator $T$ on the operators $d_{\bs k \mu}$ in the main text is given by
 \begin{gather}
 \mathscr{\hat{U}} (\bs k)
 =
  \begin{bmatrix}
  \left( g_{11}^*(\bs k)  g_{21}(-\bs k)+g_{12}^*(\bs k)  g_{22}(-\bs k) \right) &
\left( -g_{11}^*(\bs k)  g_{11}(-\bs k)-g_{12}^*(\bs k)  g_{12}(-\bs k) \right) \\
   \left( g_{21}^*(\bs k)  g_{21}(-\bs k)+g_{22}^*(\bs k)  g_{22}(-\bs k) \right) &
\left( -g_{21}^*(\bs k)  g_{11}(-\bs k)-g_{22}^*(\bs k)  g_{12}(-\bs k) \right) 
   \end{bmatrix}.
\end{gather}

\subsection{B. Projected Hamiltonian in the Wannier basis} 
In this section, we derive the effective model in the WO basis. 
We start from the extended Hubbard model with Coulomb interactions on the kagome lattice.
\beq \nonumber
\mathscr{H} &=&\mathscr{H}_0 +\mathscr{H}_I\\ 
\mathscr{H}_0 &=&  \sum_{ i j  \alpha \beta } c_{i \alpha }^{\dagger} T_{ij,\alpha\beta}c_{j \beta } \\
\mathscr{H}_I &=& \sum_{i j \alpha \beta } U_{j,\alpha \beta}n_{i \alpha }n_{i+j \beta } \quad .
\label{KagomeSOC_appd}
\eeq
The hopping matrix can be diagonalized by $V_k^\dag$ with $V_k T_k V_k^\dag = \text{diag}(\epsilon_{k,1},\epsilon_{k,2},...) $. We then introduce the band basis $d_{\bm{k}} = V_{\bs{k}}c_{\bs{k}}$ and projection operator $P = \prod_{\bs{k},\mu>2} (1-d^\dag_{\bs{k}\mu}d^\dag_{\bs{k}\mu})$.  It's not difficult to show $P d_{\bs{k}\mu} P = d_{\bs{k}\mu}$ for $\mu=1,2$ and $Pd_{\bs{k}\mu}P=0$ for $\mu>2$. The projection operator would only keep the lowest two bands and project out the high energy degrees of freedom. The effective Hamiltonian is defined as 
\begin{equation}
    H = P\mathscr{H}P
\end{equation}

Since the lowest two bands have non-trivial topology, the Hamiltonian in the band basis contains long-range interactions. In order to derive a local Hamiltonian with exponentially decayed interactions, we introduce the WOs as described in previous section $b_{\bs{k}} =U_{H,\bs{k}}d_{\bs{k}}$. In the WO basis, it's not difficult to show: 
\begin{eqnarray}
    Pc_{\bs{k}\mu} P& =&  \sum_\mu (V_{\bs{k}}^\dag U_{H.\bs{k}}^\dag)_{\mu\nu} b_{\bs{k}\nu} 
\end{eqnarray}
Then we can replace all the terms in the Hamiltonian with their projected formula and then the effective Hamiltonian reads 
\beq
H &=& H_0 + H_I \\ \nonumber
H_0 &=& \sum_{\substack{ij \\ \mu \nu}} b_{i \mu}^\dag t_{ij}^{\mu \nu} b_{j \nu}  \\ \nonumber
 H_{I} &=& \sum_{\substack{ijkl\\\mu\mu'\nu\nu'}} u_{\mu\mu'\nu\nu'}(j,k,l) b_{i\mu}^\dag b_{i+j\mu'}
 b_{i+k\nu}^\dag b_{i+l\nu'} 
\eeq
where 
\beq \label{eq:int_formula}
t_{ij}^{\mu\nu}& =&\sum_{\mu'\nu'i'j'} T_{i'j'}^{\mu'\nu'}w_{\mu'\mu,i'-i}^*w_{\nu'\nu,j'-j} \nonumber \\
u_{\mu\nu\mu'\nu'}(j,k,l)&=&\sum_{ {e,i, \alpha\gamma}} 
U_{e,\alpha\gamma} w_{\alpha\mu,i}^* w_{\alpha\nu,i-j}w_{\gamma\mu',i-k+e}^*w_{\gamma\nu',i-l+e}  \quad ,
\eeq 
and $w_{\alpha\nu,i} =\frac{1}{N} \sum_k (V_{\bs{k}}^\dag U_{H,\bs{k}}^\dag)_{\mu\nu}e^{i\bs{k}\cdot \bs{r}_i}$ denotes the real space wave functions of WOs. 

\begin{figure}[t!]
\includegraphics[width=4.2in,height=2.5in]{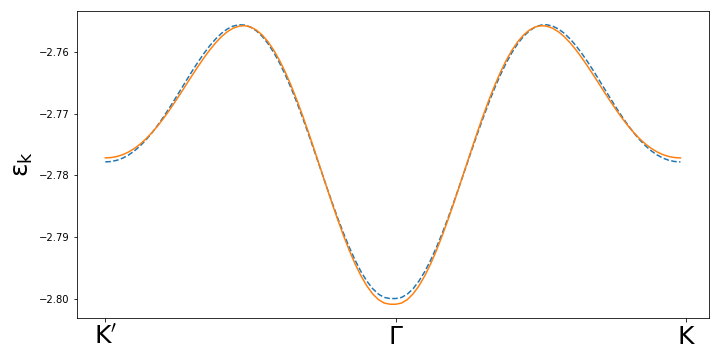}
\caption{A comparison of dispersions along high symmetry lines of the lowest degenerate doublet of kagome bands. Dashed line is the original kagome band and solid line is from the effective model with Wannier orbitals. }\label{BSComparison}
\end{figure}

The hopping parameters of the original kagome lattice are: $t_1=1,t_2=-0.3,\lambda_1=0.28,\lambda_2=0.2$. The effective hopping takes form of $t_{ij}^{\mu\nu} = \delta_{\mu\nu} t_{j-i}$. We truncate to the third-neighbor hopping and list the hopping strength in Table~\ref{tab:hop_wo}. In Fig.~\ref{BSComparison}, we compare the original band structures with the dispersion generated by Table~\ref{tab:hop_wo} and a good consistency is seen.
\begin{table}[]
    \centering
    \begin{tabular}{c|c | c | c | c | c | c }
        \hline 
         $n,m$ & (0,1) & (1,0) & (0,-1) & (-1,0) & (-1,-1) & (1,1) \\ 
         \hline 
         $t_{n\bs n_1+m\bs n_2}$  &  -0.000698& -0.000698& -0.000698&  -0.000698& -0.000698& -0.000698 \\ 
         \hline 
         $n,m$ & (-2,0) & (2,0) & (0,2) & (0,-2) & (-2,-2) & (2,2) \\ 
         \hline 
         $t_{n\bs n_1+m\bs n_2}$  &  -0.001945&    -0.001945&   -0.001945&   -0.001945 &    -0.001945&   -0.001945 \\ 
         \hline 
          $n,m$ & (-2,-1) & (-1,-2) & (-1,1) & (1,-1) & (1,2) & (2,1) \\ 
         \hline 
         $t_{n\bs n_1+m\bs n_2}$  & -0.003026 &  -0.003026 & -0.003026 &  -0.003026 & -0.003026 &  -0.003026 \\ 
         \hline 
    \end{tabular}
    \caption{Hopping parameters in the WO basis}
    \label{tab:hop_wo}
\end{table}

For  the interaction term, we consider the onsite-Hubbard interactions $U$, the nearest-neighbor Coulomb interactions $V$ and next-nearest-neighbor Coulomb interactions $V'$ in the original kagome models. These would generate various interaction terms in the effective model. The numerical values of each term are shown in Table~\ref{tab:int_wo_I} and Table~\ref{tab:int_wo_II}.  For the terms generated by $U$, we omit ones that are less than $10\%$ of the leading-order contributions. For the terms generated by $V,V'$, we only keep the leading order contributions, The sub-leading part is about $30\%$ of the leading-order contributions. Even though various types of interactions could be induced, all the terms in the Hamiltonian are exponentially decayed.
\begin{table}[] 
\centering
\begin{tabular}{ c | c | c }
\hline
   $ m,n $  &  $ \mu,\nu $  &  $ u_{\mu\mu\nu\nu}(0,m \bs n_1 + n \bs n_2 ,m \bs n_1 + n \bs n_2)$ \\
     \hline 
     0,0 & 0,1 &$ 0.078U+0.320V + 0.146V'$\\
     \hline
     -1,0 & 0,1 & $ 0.078U+0.320V + 0.146V'$\\
     \hline
     0,1 & 0,1 & $ 0.078U+0.320V + 0.146V'$\\
      \hline
     -1,-1 & 0,1 & $0.033V+ 0.142V'$ \\
      \hline
     1,1 & 0,1 & $0.033V+ 0.142V'$ \\
      \hline
     -1,1& 0,1 & $0.033V+ 0.142V'$ \\
      \hline
     1,0& 0,0 &$ 0.066V+0.142V'$\\
      \hline
     0,1& 0,0 & $0.066V+0.142V'$\\
      \hline
     1,1&0,0 & $0.066V+0.142V'$\\
     \hline 
      1,0& 1,1 &$ 0.066V+0.142V'$\\
      \hline
     0,1& 1,1 & $0.066V+0.142V'$\\
      \hline
     1,1&1,1& $0.066V+0.142V'$\\
     \hline
\end{tabular}
  \caption{Interaction terms in the WO basis (part I).}
  \label{tab:int_wo_I}
\end{table}

\begin{table}[]
    \centering
    \begin{tabular}{ c | c | c | c | c | c  }
    \hline 
         $ m,n,k,l $ &  $\mu,\nu$ &  $ u_{\mu\mu\nu\nu}(0,0,m \bs a_1 + n \bs n_2 ,k \bs a_1 + l \bs n_2)$ &  $ m,n,k,l $ &  $\mu,\nu$ &  $ u_{\mu\mu\nu\nu}(0,0,m \bs a_1 + n \bs n_2 ,k \bs a_1 + l \bs n_2)$\\
          \hline 
        0,1,0,0 & 0,1 &$ 0.02U$ & 1,0,0,0 & 0,1 & $0.016U$\\
        \hline 
        0,0,0,-1 & 1,0 &$ 0.02U$ &0,0,-1,0 & 1,0 &$0.016U$ \\
        \hline 
        0,-1,0,0 & 1,0 &$ 0.02U$ &-1,0,0,0 & 1,0 & $0.016U$ \\
        \hline 
        0,0,0,1 & 1,0 & $ 0.02U$ & 0,0,1,0 & 1,0 &$0.016U$\\
        \hline 
        0,0,0,1 & 0,1 & $ 0.02U$ & 0,0,1.0 & 0,1 &$0.016U$\\
        \hline 
        0,1,0,0 & 0,1 & $ 0.02U$ &1.0,0,0 & 0,1 &$0.016U$\\
        \hline 
        0,0,0,1 & 0,1 & $ 0.02U$ &  0,0,1,0 & 0,1 &$0.016U$\\
        \hline 
        0,-1,0,0 & 1,0 & $ 0.02U$ & -1,0,0,0 & 1,0 & $0.016U$\\
        \hline 
        1,0,0,1 & 1,0 & $0.018U$ & 0,1,1,0 & 0,1 & $0.018U$ \\
        \hline 
        0,-1,1,0 & 1,0 & $0.018U$ & -1,0,0,1 & 0,1 & $0.018U$ \\
        \hline 
        1,0,0,-1 & 1,0 & $0.018U$ & 0,1,-1,0 & 0,1 & $0.018U$ \\
        \hline 
        0,-1,1,0 & 1,0 & $0.018U$ & -1,0,0,1& 0,1 & $0.018U$ \\ 
        \hline 
        \end{tabular}
    \caption{Interaction terms in the WO basis (part II).}
    \label{tab:int_wo_II}
\end{table}

\subsection{C. Real space Wannier functions and approximate Hamiltonian} 
\begin{figure}
    \centering
    \includegraphics[scale=0.3]{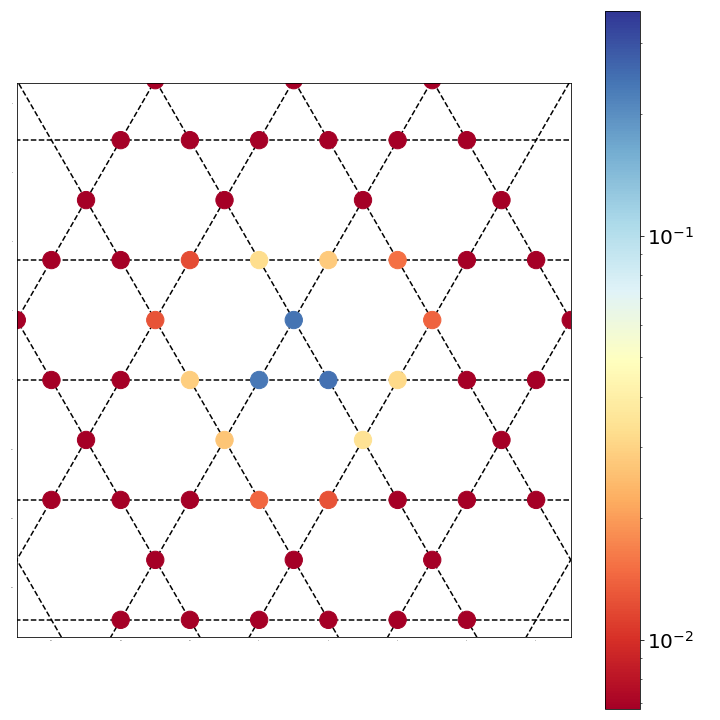}
    \includegraphics[scale=0.3]{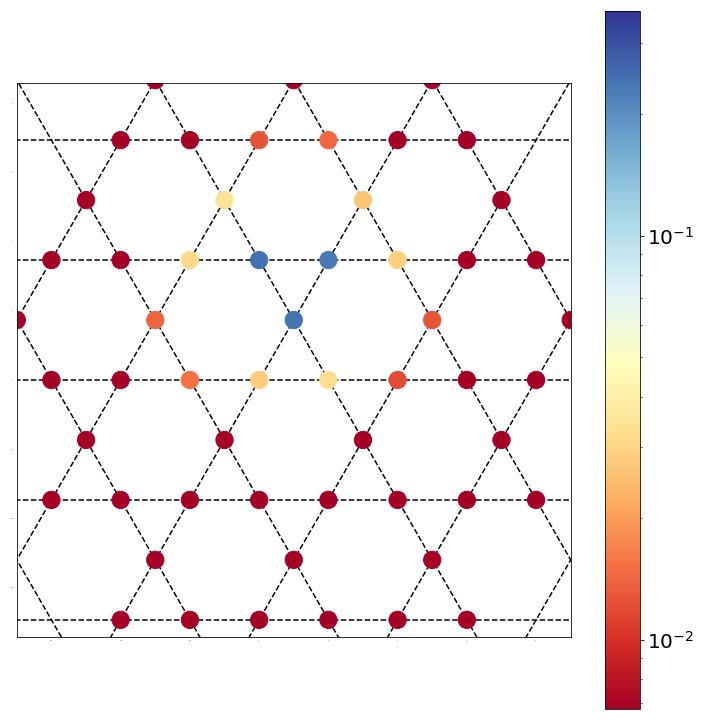}
    \caption{Probability distributions of two maximally localized WOs.}
    \label{fig:loc_wo}
\end{figure}
We now turn to the real-space pattern of WOs and the approximate Hamiltonian. 
After constructing the WOs from the smooth gauge, we also use the standard maximal-localization procedure to make the Wannier orbitals even more localized. The final amplitude of two WOs are shown in Fig.~\ref{fig:loc_wo} 

Approximately, the leading contributions of each WO come from three sites (marked as blue points in Fig.~\ref{fig:loc_wo}). It allows us to construct the approximate formula of WOs
\beq
&b_{r,1}^\dag& \sim A(c_{r,a+}^\dag +c_{r,b+}^\dag +c_{r,c+}^\dag ) \nonumber \\
&b_{r,2}^\dag &\sim A(c_{r+\bs{n}_1,a-}^\dag +c_{r-\bs{n}_2,b-}^\dag +  c_{r,c-}^\dag ) \label{eq:wf_approx}
\eeq 
Here we are using the eigenstates of $S^x$: $|\pm\rangle=\frac{1}{\sqrt{2}}|\uparrow \pm \downarrow\rangle $. $a,b,c$ denotes three sub-lattices located at $(-1/4,\sqrt{3}/4),(1/4,\sqrt{3}/4),(0,\sqrt{3}/2)$. Numerically, we find $A\sim 0.5$ and the sub-leading contributions from each site is less than $10\%$.

We want to emphasize that the two WOs don't have a common center in the topological non-trivial case (even after the maximal-localization procedure).  The sum of two Wannier centers $\bar{r}$ is a gauge-invariant quantity and won't change during the localization procedure. It lies at the center of the hexagon (setting one center of the hexagon as the origin) in the topological non-trivial case. If we maximally localized one of the Wannier orbitals, its Wannier center lies at the center of the triangular. Consequently, the center of another Wannier orbitals has to deviate from the first one in order to make $\bar{r}$ stay at the center of the hexagon.  However, in a topological  trivial
 case, we can always make two WOs be Kramers doublets; in turn,
the  two WOs have a common center due to the time-reversal symmetry.

According to the approximate formula in Eq.~\ref{eq:wf_approx}, we can also write the approximate interactions as:
\beq 
H_{I} &=& \sum_{ij} (U+4V+4V')A^4b_{1i}^\dag b_{1i} (\delta_{j,i}+\delta_{j,i-\bm{n}_1} + \delta_{j,i+\bm{n}_2}) b_{2j}^\dag b_{2j}
\nonumber \\
&+&\sum_{ij,\mu} (V+2V')A^4b_{\mu i}^\dag b_{\mu i} (\delta_{j,i+\bm{n}_1}+\delta_{j,i+\bm{n}_2} + \delta_{j,i+\bm{n}_1+\bm{n}_2}) b_{\mu j}^\dag b_{\mu j} 
\nonumber \\
&+&\sum_{ij,\mu} 2V'A^4b_{1 i}^\dag b_{1 i} (\delta_{j,i+\bm{n}_1+\bm{n}_2}+\delta_{j,i-\bm{n}_1-\bm{n}_2} + \delta_{j,i-\bm{n}_1+\bm{n}_2}) b_{2j}^\dag b_{2j} 
\label{eq:approx_int}
\eeq 
The above expressions are also consistent with the leading contributions from our numerical results shown in Table~\ref{tab:int_wo_I} and Table~\ref{tab:int_wo_II}.

We now discuss the effect of sub-leading terms that are not included in Eq.~\ref{eq:approx_int}. We divide the sub-leading interactions into two parts: density-density and normalized hopping. The density-density part takes the form of $v_{ij,\mu\nu}n_{i\mu}n_{j\nu}$. We found such terms to favor the same order as the leading interactions, This part would only modify the phase boundaries. The normalized hopping term can be written as $V_{ijk,\mu\nu\alpha}n_{i\mu}d_{j\nu}^\dag d_{k\alpha}$. At the 
mean-field level, we can decouple this term as $V_{ijk,\mu\nu\alpha} (\langle n_{i\mu}\rangle d_{j\nu}^\dag d_{k\alpha}+ n_{i\mu}\langle d_{j\nu}^\dag d_{k\alpha}\rangle )$. It would normalize the hopping strength and favors some bond orders. However, the average strength of the normalized hopping term is about $80\%$ smaller than the dominant terms. In summary, the approximate interactions in Eq.~\ref{eq:approx_int} already faithfully describe the original kagome system.

\subsection{D. Entwined order parameters} 
\begin{figure}[h!]
\includegraphics[width=3.5in]{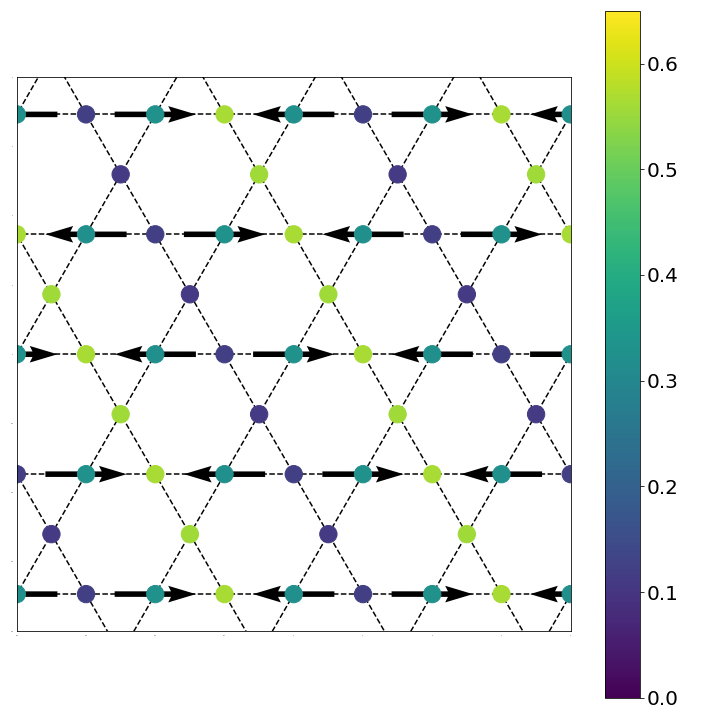}
\caption{Real space pattern of the DW$_{\text{t}}$.  The color represents the filling of each site and
the arrow denotes the magnitude and direction of in-plane local moment.
 We can observe a collinear order in the charge channel that is decorated by a collinear order 
 in the spin channel.}
 \label{fig:real_space_pattern}
\end{figure}

In this section, we discuss the connections between order parameters in WO basis and kagome basis. The charge and spin order parameters along $x$ direction in the kagome basis can be written as 
\beq 
\phi_q^c&=& \langle Pc^\dag_k c_{k+q} P\rangle 
\sim  \langle\sum_\alpha w_{\alpha\nu k} w^*_{\alpha\mu k+q} b_{k\nu}^\dag b_{k+q\mu} \rangle \nonumber \\
\phi_q^x&=& \langle Pc^\dag_{k}\sigma^x c_{k+q} P\rangle
\sim  \langle\sum_\alpha \sigma^x_{\alpha\gamma} w_{\alpha\nu k} w^*_{\gamma\mu k+q} b_{k\nu}^\dag b_{k+q\mu} \rangle
\eeq 
Combining the above expressions with Eq.~\ref{eq:wf_approx}, we find the following approximate formula as shown in the main text
\beq 
\phi_q^c& \sim &  
 \frac{4+e^{iq\bs{n}_1}+e^{-iq\bs{n}_2} }{2}A^2(n_{1,q} +n_{2,q} ) + 
 \frac{2-e^{iq\bs{n}_1}-e^{-iq\bs{n}_2} }{2}A^2(n_{1,q} -n_{2,q} ) \nonumber \\
 \phi_q^x &\sim &
 \frac{2-e^{iq\bs{n}_1}-e^{-iq\bs{n}_2} }{2}A^2(n_{1,q} +n_{2,q} ) + 
 \frac{4+e^{iq\bs{n}_1}+e^{-iq\bs{n}_2} }{2}A^2(n_{1,q} -n_{2,q} ) 
\eeq 
This suggests a single charge order (or spin order) in the WO basis produces entwined  charge and spin orders in the 
kagome basis. In Fig.~\ref{fig:real_space_pattern}, we illustrate this by plotting the real-space pattern of the 
DW$_\text{t}$ phase (charge order in the WO basis). 

\subsection{E. U(1) Slave-spin method} 
In the $U(1)$ slave spin method, we rewrite the electron operators as
$b_{i\mu} =S^{-}_{i\mu} f_{i\mu }$
 where $f$ is the fermion operator and $S^-$ is the spin operator.  The original Hilbert space can be mapped onto an extended space of slave spins and $f$ fermions according to the following correspondence:
\begin{eqnarray}
&&|1\rangle_b = |1\rangle_f |\uparrow\rangle_{S} \nonumber \\
&&|0\rangle_b = |0\rangle_f |\downarrow\rangle_{S}
%\nonumber 
\end{eqnarray}
To project out the unphysical degree of freedom, we enforce the constraint:
 $S^{z}_{i\mu}+\frac{1}{2} =f_{i\mu}^\dag f_{i\mu }$  
 
The slave spins carry the $U(1)$ charge which allows us to rewrite the interactions as  
\begin{eqnarray} 
b_{i\mu}^\dag b_{i\mu} b_{j\mu'}b_{j\mu'} = (S_{i\mu}^z +\frac{1}{2})(S_{j\mu'}^z +\frac{1}{2})
\end{eqnarray} 
For the non-local interactions with $i\ne j$, we further decouple it via 
\begin{equation}
(S_{i\mu}^z +\frac{1}{2})(S_{j\mu'}^z +\frac{1}{2}) \sim -
\langle S_{i\mu}^z +\frac{1}{2}\rangle \langle S_{j\mu'}^z +\frac{1}{2}\rangle + 
\langle S_{i\mu}^z +\frac{1}{2}\rangle (S_{j\mu'}^z +\frac{1}{2}) + 
(S_{i\mu}^z +\frac{1}{2}) \langle S_{j\mu'}^z +\frac{1}{2}\rangle 
\end{equation}

Then the effective Hamiltonian is 
\begin{eqnarray}
&H_f& = \sum_{ij} t_{ij}^{\mu\mu'} \eta_{ij} f_{i\mu}^\dag f_{j\mu'} + \sum_i \lambda_{i\mu}  f_{i\mu}^\dag f_{i\mu} \\
&H_S& = \sum_{ij,\mu\mu'} t_{ij}^{\mu\mu'} (\chi_{ij}^{\mu\mu'} \langle z^\dag_{i\mu} \rangle z_{j\mu'} +\text{h.c.})-\sum_{i,\mu} \lambda_{i\mu} S_{i\mu}^z + \sum_i u_{1122}(0,0,0)(S_{i,1}^z+\frac{1}{2})(S_{i,2}^z+\frac{1}{2} ) + \sum_{i\mu} m_{i\mu} (S_{i\mu}^z+\frac{1}{2}) \nonumber
\end{eqnarray}
where $\lambda_\mu$ is the Lagrangian multiplier and operators $z_{i\mu} = \frac{1}{\sqrt{1/2+S_{i\mu}^z}}S^-_{i\mu}\frac{1}{\sqrt{1/2-S_{i\mu}^z}} $ are introduced to obtain the correct non-interacting limit. Other parameters are determined self-consistently by 
\begin{eqnarray}
&&\eta_{ij}^{\mu\mu'} = \langle z_{i\mu}^\dag\rangle_{H_S} \langle z_{j\mu'}\rangle_{H_S} \nonumber \\
&&  \chi_{ij}^{\mu\mu'} = \langle f_{i\mu}^\dag f_{j\mu'}\rangle_{H_f} \nonumber \\
&& m_{i\mu} = \sum_{j,\nu}u_{\mu\mu\nu\nu}(0,j,j) \langle S^z_{i+j\nu}+\frac{1}{2}\rangle_{H_S}
\end{eqnarray}  

Effectively, our slave-spin method introduces a wave-function Ansatz: $P_G|HF\rangle$ and variationally find the state with the lowest energy. $P_G$ is a Gutwiller projector that disfavors double occupancy and $|HF\rangle$ denotes the ground state of a free electrons model with Hartree-Fock contributions.  

At each $(U,V')$, we randomly initialized 15 sets of variables $(\eta_{ij}^{\mu\mu'},\chi_{ij}^{\mu\mu'},m_{i\mu},\lambda_{i\mu})$ and find self-consistent solutions of each initialization. Any type of three-sublattice and two-sublattice orders are allowed in the calculations, and other types of orders are not favored by interactions.
The final ground state corresponds to the solution with the lowest free energy. In addition, we set the temperature to be $\beta W=1000$.

\end{document}